\documentclass[sigconf]{acmart}
\AtBeginDocument{%
  \providecommand\BibTeX{{%
    \normalfont B\kern-0.5em{\scshape i\kern-0.25em b}\kern-0.8em\TeX}}}

\setcopyright{acmlicensed}
\copyrightyear{2018}
\acmYear{2018}
\acmDOI{XXXXXXX.XXXXXXX}

\acmConference[Conference acronym 'XX]{Make sure to enter the correct
  conference title from your rights confirmation emai}{June 03--05,
  2018}{Woodstock, NY}
\acmISBN{978-1-4503-XXXX-X/18/06}

\usepackage{graphicx}
\usepackage{hyperref}

\usepackage{amsfonts,amssymb}
\usepackage{pythonhighlight}
\usepackage{array}
\usepackage{amsmath}

\PassOptionsToPackage{prologue, dvipsnames}{xcolor}
\usepackage{caption}
\usepackage[dvipsnames]{xcolor}  % 更全的色系
\usepackage{listings}  % 排代码用的宏包

\begin{document}

\title{Anywhere: A Web Crawler Automation Management Interface}

\author{Jinwei Lin}
\authornote{First author of this research.}
\email{Jinwei.Lin@monash.edu}
\orcid{0000-0003-0558-6699}
\affiliation{%
  \institution{Monash University}
  \city{Shenzhen}
  \state{Guangdong}
  \country{China}
}


\begin{abstract}
Web crawling projects or design is significant in the current information age. Using the web spider or crawler can automatically search and collect a huge amount of internet information. As one of the most popular web crawler frameworks, Scrapy is robust in abundant functions but weak in easy operation. In this paper, we provide a framework Anywhere, for optimising the usage feeling and improving the use efficiency of the web crawling management of Scrapy. We analyse the whole workflow of a web crawling project of Scrapy and design two main functions in Anywhere, one is quickly generating a Scrapy project with the preset temperatures, the other is repeatable configuration function for the Scrapy project setting. Beside, with Anywhere, users can easily directly manage multiple Scrapy projects with a file folders architecture. Compared with normal Scrapy project interactive coding development, we test Anywhere with enough experiments that show Anywhere can improve the development efficiency of Scrapy projects to about 200\%. For the multiple project management in code interaction level, the developing efficiency is improved to about 300\%. We simplify the procedure to quickly generate a simple spider project with Scrapy. Anywhere can assist the development of Scrapy is useful for the design of large batch  concurrent projects at coding level.
\end{abstract}






\received{20 February 2007}
\received[revised]{12 March 2009}
\received[accepted]{5 June 2009}

\maketitle

\section{Introduction}
ACM's consolidated article template, introduced in 2017, provides a
consistent \LaTeX\ style for use across ACM publications, and
incorporates accessibility and metadata-extraction functionality
necessary for future Digital Library endeavors. Numerous ACM and
SIG-specific \LaTeX\ templates have been examined, and their unique
features incorporated into this single new template.

% >>:=======================================================================================
\section{Introduction}

Web crawler is an important tool or method to collect huge amounts of information from the internet, which is applied in many implementation fields, such as analytical finance, natural language processing (NLP), software engineering and development \cite{singrodia2019review}. Web crawler is an automatic information collecting robot program that can implement the work of web information crawling and collection. To improve the efficiency of making use of the web scraping, many crawler program toolkits or frameworks are designed. For the Python world, the Scrapy framework is one of the most popular web crawling frameworks \cite{kaiying2020optimisation}. The application method to use the Scrapy framework is coding in a python file to recall Scrapy or using the terminal or Command (CMD) interface to make the direct interaction with the Scrapy framework.

Although, this kind of directly interacting with the terminal coding is direct control but can only code one common at a time \cite{persson2019evaluating}, which means this interaction method is low in efficiency. Moreover, the application method of pure terminal coding is not easy to use, which will increase the user barriers and difficulties. Compared with the graphic interaction with the program, the direct coding method won the advantage of more lower level operations but lost more convenience in usage \cite{jansen1998graphical}. But for those applications or frameworks which focus more on the source code level interaction and more degree of freedom in source code interaction, the interaction method of source code second programming or coding in terminal will be more available. Scrapy is a web crawling framework that focuses on terminal interaction and code second development \cite{el2020using}, terminal interaction is the first recommended application and re-development method for Scrapy projects.

Nevertheless, terminal direct interaction will increase the controllable configuration and programming for a Scrapy project, but this kind of project design structure will cause two main kinds of bad influences, one is the difficulty of usage operation, which can be address by using the visual project manage methods to make the operations on a GUI interface for a web crawler application \cite{dobrokvashina2021improved}. This is a feasible research direction and will be easy and friendly for most of the users who have little relative usage experiences. The second is improving the processing algorithm of Scrapy to increase the crawling speed \cite{kaiying2020optimisation}, or improving the interactive abilities and functions of APIs to improve the easy use of Scrapy and robust intelligent operation abilities with other programs. Not that  these two methods are not opposite but complement each other. Focusing on the GUI application will research more on the GUI interaction, focusing on the code APIs improving will research more on the code level improvement and updation. Our work toolkit with the improvement for Scrapy framework is represented as Anywhere, which means crawling anywhere.

% Anywhere will focus on Scrapy framework but not change its source code, which means Anywhere is individual from Scrapy and the design idea and solution can be used in other application fields. The relative experiment code is open source in GitHub:\href{https://github.com/JYLinOK/anywhere}{https://github.com/JYLinOK/anywhere}.

Anywhere will focus on Scrapy framework but not change its source code, which means Anywhere is individual from Scrapy and the design idea and solution can be used in other application fields. The relative experiment code is open source in GitHub, and will be provided after the paper acceptation.
% <<:=======================================================================================

% >>:=======================================================================================
\section{Literature Review}

In this section, we will make a relative introduction for the literature review of using Scrapy and the background of our research, to state the significance of the research of Anywhere.

% >>:-----------------------------------------------------------------------------------
\subsection{Difficulties in using Scrapy}
The normal application or development method of Scrapy is based on terminal interaction coding. Therefore, for easy to use, Scrapy has its own special technical difficulty \cite{kouzis2016learning}.  Except for the interaction methods of pure coding in terminal or CMD and modifying the source code of Scrapy, another method to use Scrapy is to make an encapsulation for the APIs of Scrapy and design a GUI application \cite{rice2023developing}. Compared with designing a GUI application for Scrapy \cite{wang2023easyspider}, using the APIs to make the code level development for Scrapy general application will have more degree of freedom in more detailed control and functions design. Because Scrapy is developed based on the Python programming language \cite{myers2015choosing}, the recommended method to implement the encapsulation development for Scrapy is using Python as the developing language.  Whatismore, the actual procedures of the method to design  a web crawling project is difficult to reuse in other projects that have different application issues, therefore, the procedures of designing a web crawling project is difficult to be generalised. But the method idea can be generalised due to its low limitation and large applicability, which is one of our research key points.

Therefore, using Python to design or develop a manager program or toolkit for Scrapy to make a more easy to use crawler will be available. Moreover, in some special application fields, such as parallel processing or designing based on quick-developing templates, and repetitive development, which can be handled by using the Python programming. 
% <<:-----------------------------------------------------------------------------------

% >>:-----------------------------------------------------------------------------------
\subsection{Relative Web Crawling Framework}

Although as one of the most popular frameworks for Python programming, Scrapy still has some feasible competitors that are other web crawling frameworks \cite{khder2021web}, such as Nutch \cite{shafiq2020ncl} using the Java language. But compared with those crawling frameworks that are developed by Python, the amount of the crawling frameworks that are developed by other languages is low. To summary and further analyse the relative web crawling framework for Scrapy, we make a survey and statistics for the top 1,000 web spider frameworks that sorted by the liked starts number in a descending order, and deleted the mistaken searched items from them, the result is shown as Table \ref{tb1}.

\begin{table}[htb]   
\begin{center}   
\caption{Analysis of GitHub's top 1,000 star sorting items.}  
\label{tb1} 
\tabcolsep=0.06cm
\renewcommand\arraystretch{1.5}
\begin{tabular}{|c|c|c|c|c|c|c|c|c|}   
\hline   \textbf{language} & \textbf{train}  & \textbf{framework}  & \textbf{relative} & \textbf{graphic} & \textbf{concurrency} \\   

\hline Python             & 17 & 6 & 30 & 2 & 6     \\
\hline Golang             & 1 & 8 & 1 & 0 & 3       \\
\hline PHP                & 1 & 3 & 4 & 0 & 0       \\
\hline Java               & 0 & 2 & 3 & 1 & 0       \\
\hline JavaScript         & 0 & 2 & 10 & 1 & 0       \\
\hline C\#                & 0 & 2 & 0 & 0 & 0       \\

\hline   
\end{tabular}   
\end{center}   
\end{table}

The parameter $language$ means the language used to program, the parameter $train$ represents the number of projects that is used for actual training. The parameter $framework$ represents the number of projects that are designed as a framework. The parameter $relative$ represents the number of projects that is designed not as a framework but a relative toolkit or project. The parameter $graphic$ represents the number of projects that are designed with GUI user operations. The parameter $concurrency$ represents the number of projects that are designed in a distributed or high-concurrency way.

From the survey we can draw the conclusion that Python is the most popular language that is used to design web crawler projects or related projects. Golang is also used in most of the whole projects, but focuses more on the high-concurrency development, which is based on the characteristics of native concurrency of coroutines \cite{cox2017concurrency}. Due to being same as a script language and easy to use, most important, the characteristics that native support the end operation in a browser with the web page source code \cite{gyimesi2019bugsjs}, Javascript is also used in most of the whole projects, most of these projects are relative project, in other way, means the JavaScript can not support the superior operations very well. Having the most convenience in programming and design, supporting the files operations and superior data processing well, most importantly, being the native programming language of Scrapy, that is why we selected Python as the programming language and the stady direction of our research.

% <<:-----------------------------------------------------------------------------------

% >>:-----------------------------------------------------------------------------------
\subsection{Methods to Improve Scrapy Efficiency}

In this subsection, we will discuss and analyse the research methods or directions that can be used to improve the designing or processing efficiency of Scrapy. We concluded the whole method classifying as three parts. The first kind of methods to improve the code programming in normal processing or distributed high concurrent processing, which respond more on the source code or algorithm updating \cite{zhang2022deep}, moreover, it is difficult to implement and may be conflict with the main updating processing of the work of Scrapy official maintaining team. The second kind of methods is focusing on using the GUI application design to decrease the usage difficulty and purse the target that designing a web crawler project without coding \cite{hernandez2018can},  which is useful for those users who have no coding experiences but will lack of the flexibility of design and developing in coding level. The third kind of methods is to make the Scrapy framework as a third-party component of a new framework, and adding other code and algorithm \cite{yang2019design}, to make the whole workflow of traditional Scrapy project more efficient and simple, enable the new framework the ability of manage multiple Scrapy project and updating, which is the main method idea of our research.

% <<:-----------------------------------------------------------------------------------
% <<:=======================================================================================

% >>:=======================================================================================
\section{Methodology Analysis}

In this section, we will discuss and analyse the series of methods and algorithms to implement the whole design idea of the research of Anywhere.

% >>:-----------------------------------------------------------------------------------
\subsection{Research Running Architecture}
\label{cm1}

\begin{figure}[t]
	\centering
	\includegraphics[width=0.8\columnwidth]{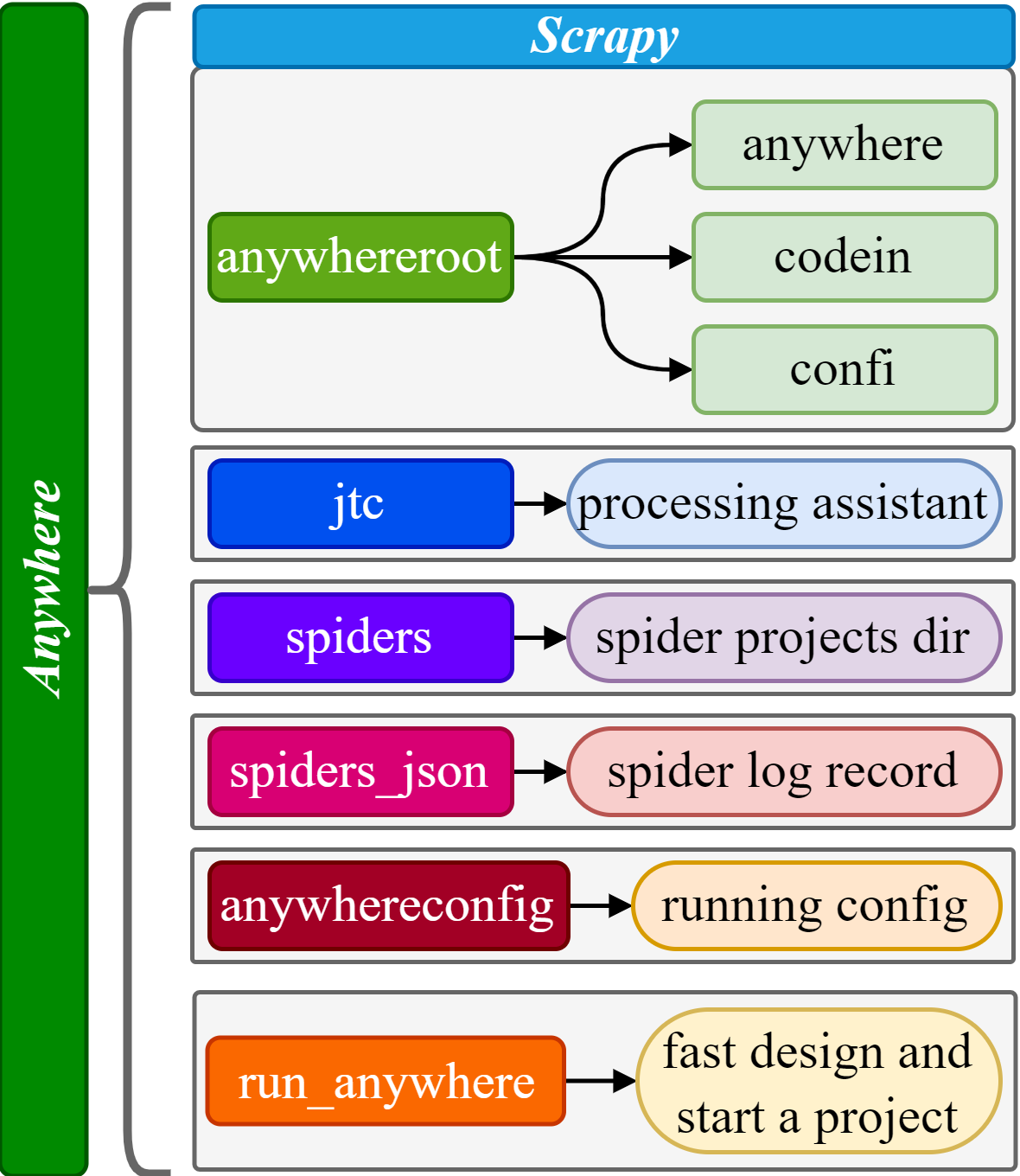}
	\caption{Running and design architecture of Anywhere.}
	\label{fig1}
\end{figure}

As shown in Figure \ref{fig1}, the whole project architecture is based on the Scrapy framework and can be divided into 6 parts: the directory $anywhereroot$ save the files which are relative to the main function of Anywhere. Among them, the $anywhere$ represents the management and controlling main file, which will call the two main function parts. The part $codein$ represents the processing module of automatically creating and generating a new Scrapy project by Anywhere. The part $confi$ represents the processing module of automatically updating the configuration of a Scrapy project that is generated by Anywhere, which is repeatable during the whole design process of the Anywhere project.

To better design and implement the whole project and research, we design a data processing toolkit library to handle the processing of the reading and writing of different file formats, such as json, csv and text. This toolkit library is open source, and the link will of the toolkit will be provided after the paper acceptation. The parameter $spiders$ represents the saving directory of the Scrapy projects that are generated by Anywhere, where each Scrapy project will be saved as its own name in an individual folder. The parameter $spiders_json$ represents the $json$ file that is used to save the log or key recording information after each Scrapy project is created by Anywhere. The parameter $anywhereconfig$ represents the configuration file that is used to config the key running parameters for Anywhere running. The parameter $run_anywhere$ represents the main running file of the whole Anywhere project, in which the users can use a few codes to quickly create and run a simple Scrapy project.

As shown in Figure \ref{fig1}, the main functions files are saved in directory $anywhereroot$, the other folders or files are the assisting components. The whole Anywhere project is designed and programmed in the file $run\_anywhere$, which will call the main function apartments from the $anywhereroot$. With directory $anywhereroot$ in the project,the framework Anywhere can be called anywhere as a library.
% <<:-----------------------------------------------------------------------------------

% >>:-----------------------------------------------------------------------------------
\subsection{Start from a Scrapy Project}
\label{cm2}

\begin{figure}[t]
	\centering
	\includegraphics[width=0.9\columnwidth]{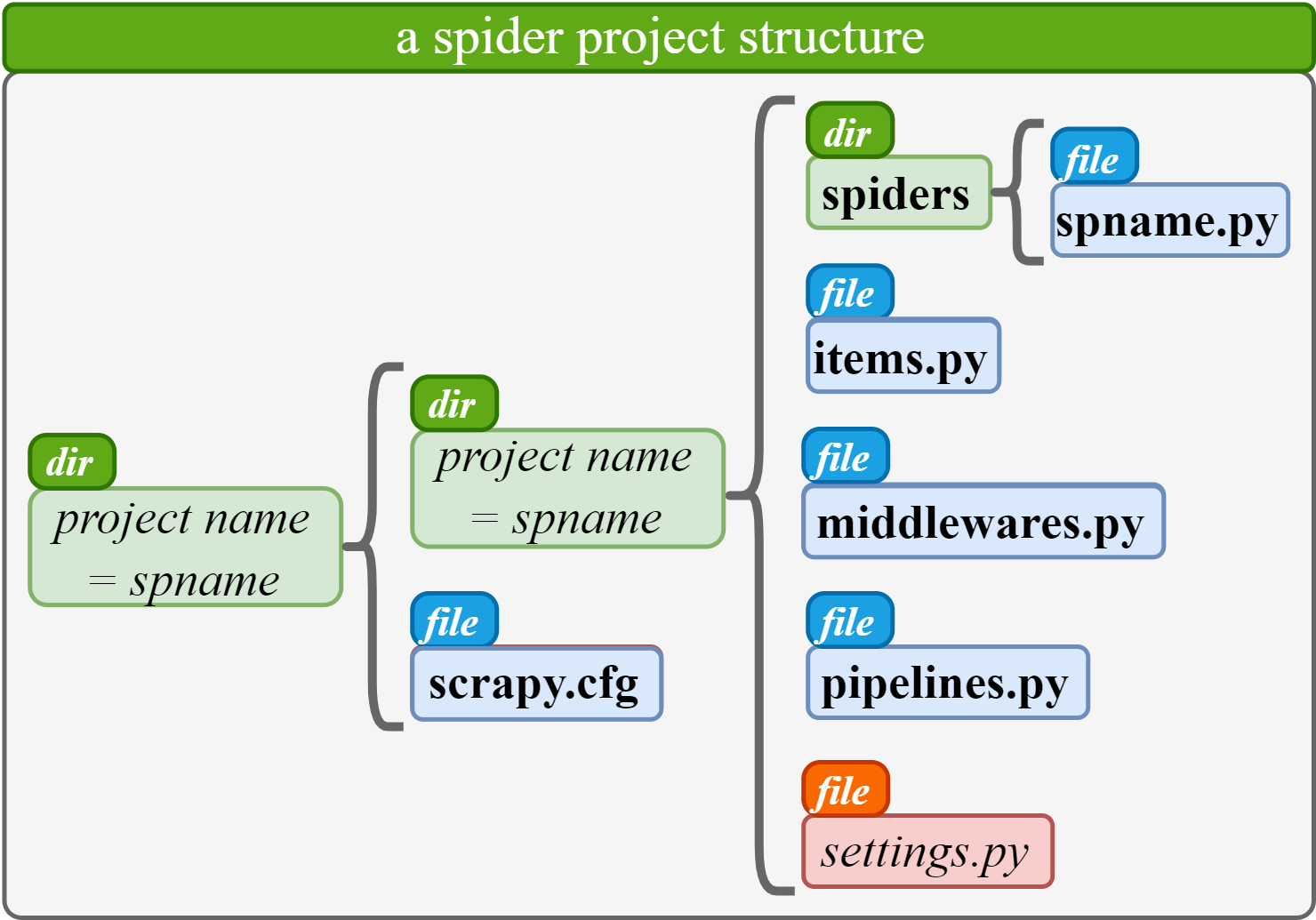}
	\caption{Structure for a classic files of a Spider project.}
	\label{fig2}
\end{figure}

Anywhere is designed and developed based on Scrapy framework, to be detailed, the further operations and process of Anywhere will be implemented after it generates a fundamental Scrapy project. As shown in Figure \ref{fig2}, when starting the processing of a project, Anywhere will generate a normal Scrapy project first, whose project structure is represented as a three directories architecture. The first directory level is the root directory of the Scrapy project, which is named with the name of the project. The second directory level is the main work directory of the Scrapy project, which is named with the name of the project. Besides, in this directory level, a configuration file of this Scrapy project will be created that is named as $scrapy.cfg$. The third directory level is the key directory of the Scrapy project, which stores the most important project files. In this directory, a folder named $spider$ will store the key file of the spider of this project, which is named as $spname.py$. Here, the name $spname$ represents the project name customised by the user. We classify those files in the second level directory $spname$ as two kinds. One is the change-once files, which can only be changed in one time, usually being generated in Anywhere generating a new Scrapy project. For example, the files: $items.py$, $middlewares.py$, $pipelines.py$, and $spname.py$ in the folder $spiders$. The other kind is repetitive change files, which can be changed multiple times in post-processing after the Anywhere generating a Scrapy project, for example, the $settings.py$ file.

% <<:-----------------------------------------------------------------------------------

% >>:-----------------------------------------------------------------------------------
\subsection{Different Processing for a Anywhere Project}
\label{cm3}

\begin{figure}[t]
	\centering
	\includegraphics[width=0.9\columnwidth]{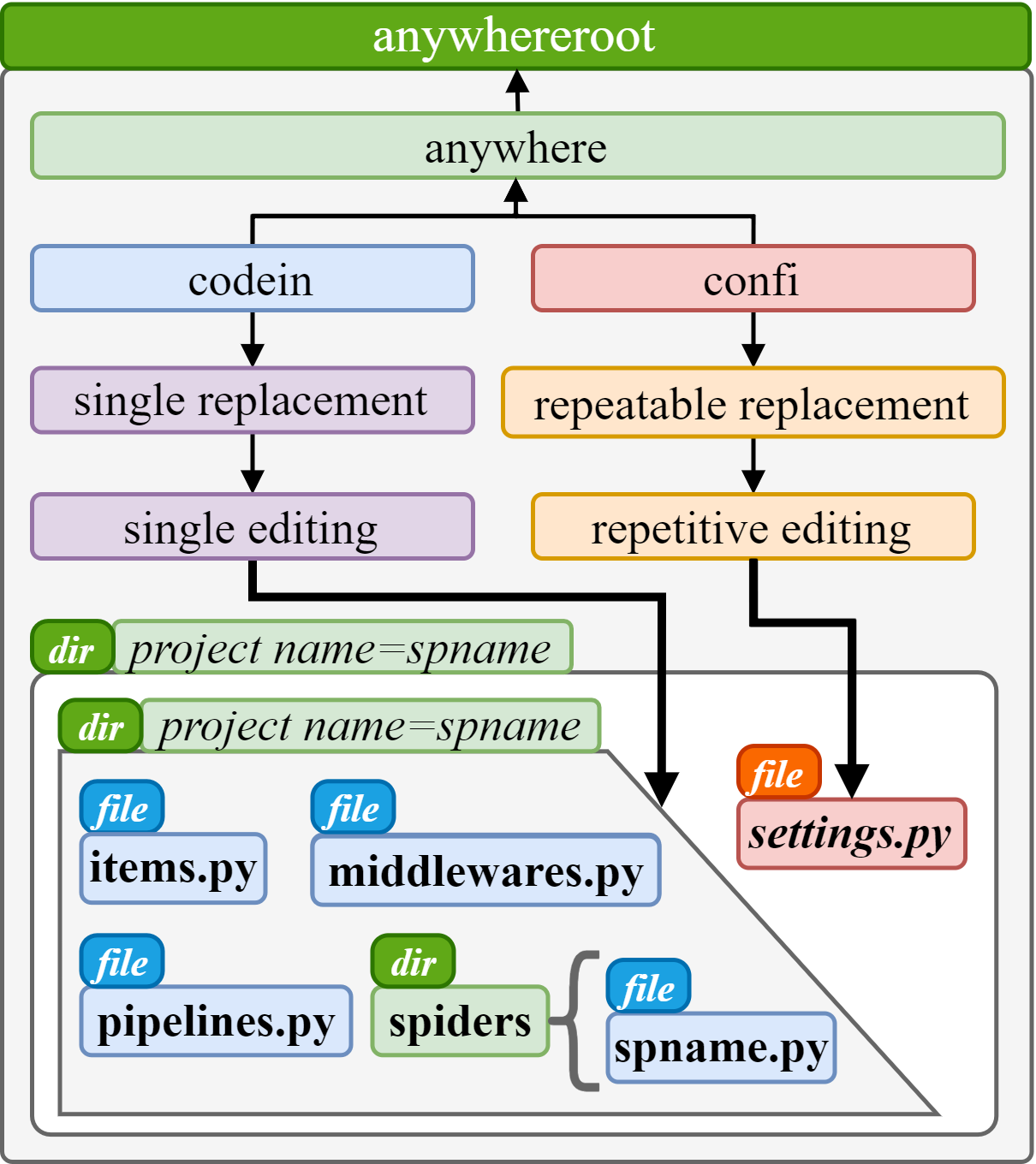}
	\caption{Different updating and processing for the files.}
	\label{fig3}
\end{figure}

The main functions of Anywhere framework focus on two directions, one is automatically updating the configuration and quickly generating a Scrapy spider project with the fixed template. Correspondingly, in our processing algorithm, there are two main processing methods, which are relative to project generating and configuration updating respectively. As shown in Figure \ref{fig3}, in the module of directory $anywhereroot$, the module $anywhere$ will call the other two modules in the same directory, one is $codein$ and the other is $confi$, which are represented by two individual Python code files. The component files in the module $condein$ are most reactive with the process of quickly generating a Scrapy spider project with the fixed template. This kind of process is a single replacement of the code of a normal Scrapy project, so it is one kind of single editing. The component files in the module $confi$ are most reactive with the process of automatically updating the configuration of the Scrapy projects. This kind of process is a repeatable replacement, so it is one kind of repetitive editing.

\begin{figure}[t]
	\centering
	\includegraphics[width=0.9\columnwidth]{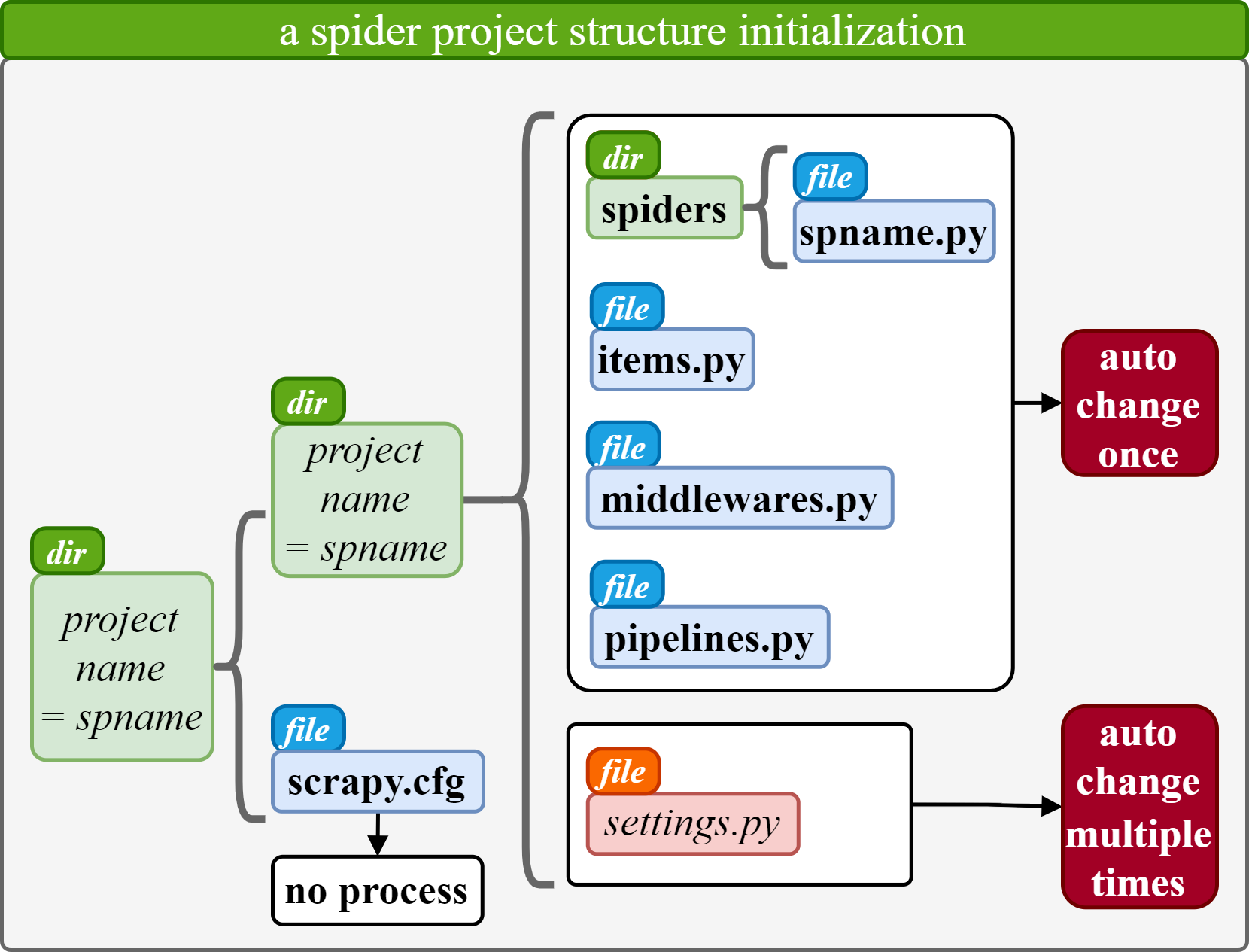}
	\caption{Initialization process of an Anywhere project.}
	\label{fig4}
\end{figure}

As shown in Figure \ref{fig4}. The change-once process will be only operated once after Anywhere framework running the relative script to generate a normal Scrapy project, which means the process of quickly generating a Scrapy spider project with the fixed template is once, and the updation and change of the Scrapy project will be implemented only after the normal Scrapy project is created. We call this process the initialization process for the module $codein$. The repetitive process can be operated at multiple times after Anywhere framework running the relative script to generate a normal Scrapy project, which means the process of automatically updating the configuration of the Scrapy projects is repetitive, and the updation and change of the configuration of Scrapy project can be implemented any time during the code running. We call this process the initialization process for the module $confi$. These are the two main module initialization processes.

\begin{figure}[t]
	\centering
	\includegraphics[width=1.0\columnwidth]{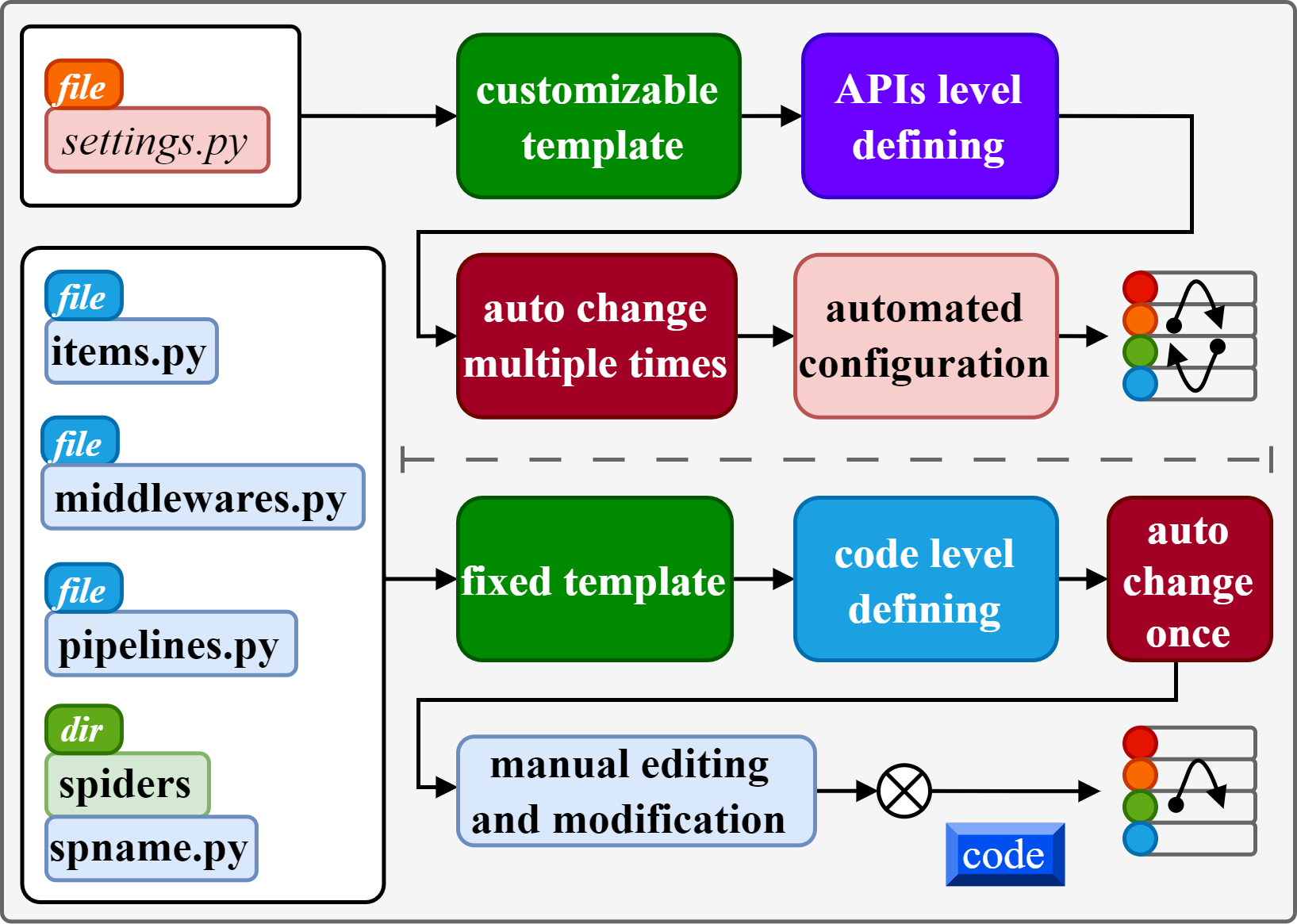}
	\caption{Different Updating Process Idea for Anywhere.}
	\label{fig5}
\end{figure}

The details of those two processes mentioned above are shown as Figure \ref{fig5}. The process about the module $Codein$, which is represented by the processes for $items.py$, $middlewares.py$, $pipelines.py$ and $spiders/spname.py$, is implemented with the preset fixed templates, and running, editing and defining in the code level, the changing of this process is once after the fundamental Scrapy project is generated by Anywhere framework. Then the user or designer can make the manual editing  or modification for the further design or operations by coding. The process about the module $confi$, which is represented by the process for $setting.py$,  is implemented with the customized templates, and running, editing and defining at the APIs level, which means these processes are handled with the APIs or functions calling the Anywhere framework.  The process of module $confi$ can be implemented multiple times, during the designing or coding time of the whole Anywhere project by the way of coding, calling and running, which means we can use this feature to design and program automated configuration functions for Anywhere framework, and this design can increase the efficiency of the repetitive configuration operations of Anywhere framework. 

Note that for now versions of the Anywhere framework, the code file $middlewares.py$ has little necessity to be updated, therefore, the process algorithm of this version of Anywhere will not implement further processes for the code file $middlewares.py$, but focusing more on other files.

% <<:-----------------------------------------------------------------------------------

% >>:-----------------------------------------------------------------------------------
\subsection{Processing Algorithm of Module Codein}
\label{cm4}

In this subsection, we will discuss and analyse the processing algorithm of the module $codein$. The key core of the process algorithm of this module is the code block searching-locating-editing algorithm. For example, when the processing algorithm wants to insert some code into a specific location of a Python file of the Scrapy project in Anywhere. First, the process algorithm will make an architecture analysis for the project. In our algorithm, a quick and efficient method to get the architecture of a Python file is required. 

\begin{figure}[t]
	\centering
	\includegraphics[width=1.0\columnwidth]{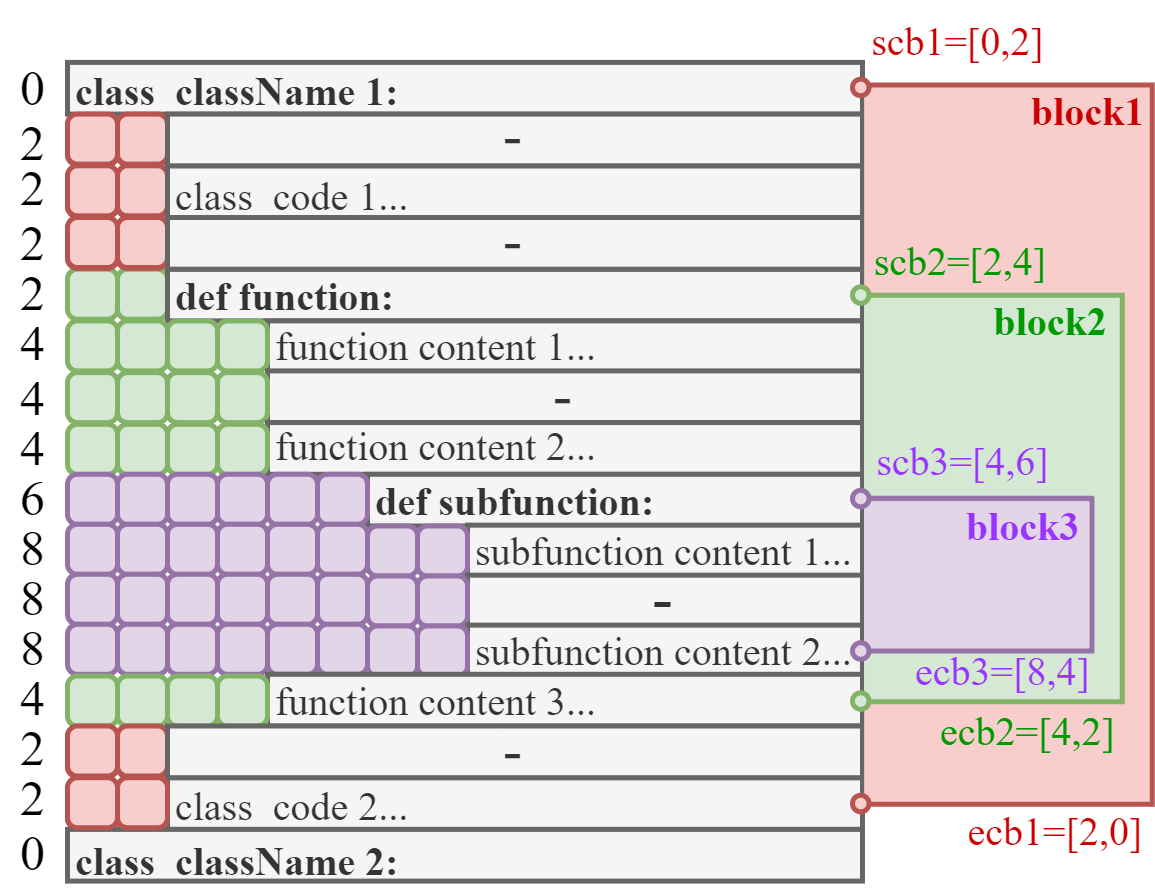}
	\caption{Code block searching and inserting algorithm.}
	\label{fig6}
\end{figure}

Therefore we design a fast search and editing algorithm for this target, which we call Code Block. As shown in Figure \ref{fig6}, for a normal Python code file, our algorithm will analyse the whole project architecture. In this analysis and processing algorithm, the tab space size in the front of a line is important. We use $t_i=n, i, n \in \mathbb{N}^+$ to represent that the number of the character size of the $NO.i$ line is $n$.

For the code that has an individual sub field, like the definition of a class, a function or Boole logic judgement statement code, from the start line to the end line of the sub field, each sub field code with an individual workspace is corresponding to a code block. As shown in Figure \ref{fig6}, we use the parameter $scb n=[f, b]$ to represent the statement of the number of the tab space of the now line and the next line of the start line of the code block.  We use the parameter $ecb n=[f, b]$ to represent the statement of the number of the tab space of the now line and the next line of the end line of the code block. In this expression, the parameter $n$ represents the number of specific code blocks in sequence, the parameter $f$ represents the number of the tab space of the now line of the specific code block,  the parameter $b$ represents the number of the tab space of the next line of the specific code block. For example, the $scb1=[0,2]$ means that the start line of code block 1 starts with 0 tab space and ends with 2 tab space, which is corresponding to the start line of the code block 1. With the same principle, the end line of the code block 1 can be represented as $ecb1=[2, 0]$. In the Figure \ref{fig6}, we use different colours to mark different code blocks. Different code blocks in a same workspace can have a nest relation. For example, code block 2 is inside the code block 1, and the code block 3 is inside the code block 2.

For the further calculation and analysis, we combine the parameters $scb$ and $efb$ to be a group as $cb$. Each line in a Python code file has its own $cb$. With the analysis of the structure of the Python code file, we can draw a conclusion or deduction that the scope or range of a code block can be gained from the analysis of the parameter $cb$ of each line. The deduction relationship is described as follows:

As shown in Equation \ref{eq1}, if $cb$ is used to judge whether the current line is a new start line of a new block. If the difference result of $f$ minus $b$ equals 0, it means the code flow will not enter a new sub code block and remain the same tab space retraction. If the difference result of $f$ minus $b$ is less than 0, it means the code flow will enter a new inner sub code block. 

\begin{equation}
\label{eq1}
cb \rightarrow scb:f-b = \left\{
\begin{aligned}
=0 \rightarrow no\ new\ start\ or\ end \\
<0 \rightarrow  -tab \rightarrow enter\ block 
\end{aligned}
\right.
\end{equation}

As shown in Equation \ref{eq2}, if $cb$ is used to judge whether the current line is the end line of the current sub code block. If the difference result of $f$ minus $b$ equals 0, it means the code flow will not enter a new sub code block and remain the same tab space retraction too. If the difference result of $f$ minus $b$ is greater than 0, it means the code flow will leave the current inner sub code block. 

\begin{equation}
\label{eq2}
cb \rightarrow ecb:f-b = \left\{
\begin{aligned}
=0 \rightarrow no\ new\ start\ or\ end \\
>0 \rightarrow +tab \rightarrow leave\ block 
\end{aligned}
\right.
\end{equation}

Combining the Equation \ref{eq1} and Equation \ref{eq2}, we can get an equation for the general code block division process for the code block in module $condein$ processing, which is represented as Equation \ref{eq3}.

\begin{equation}
\label{eq3}
cb \rightarrow f-b = \left\{
\begin{aligned}
<0 \rightarrow  -tab \rightarrow enter\ block  \\
=0 \rightarrow no\ new\ start\ or\ end \\
>0 \rightarrow +tab \rightarrow leave\ block 
\end{aligned}
\right.
\end{equation}
% <<:-----------------------------------------------------------------------------------

% >>:-----------------------------------------------------------------------------------
\subsection{Processing Algorithm of Module Confi}
\label{cm5}

\begin{figure}[t]
	\centering
	\includegraphics[width=0.9\columnwidth]{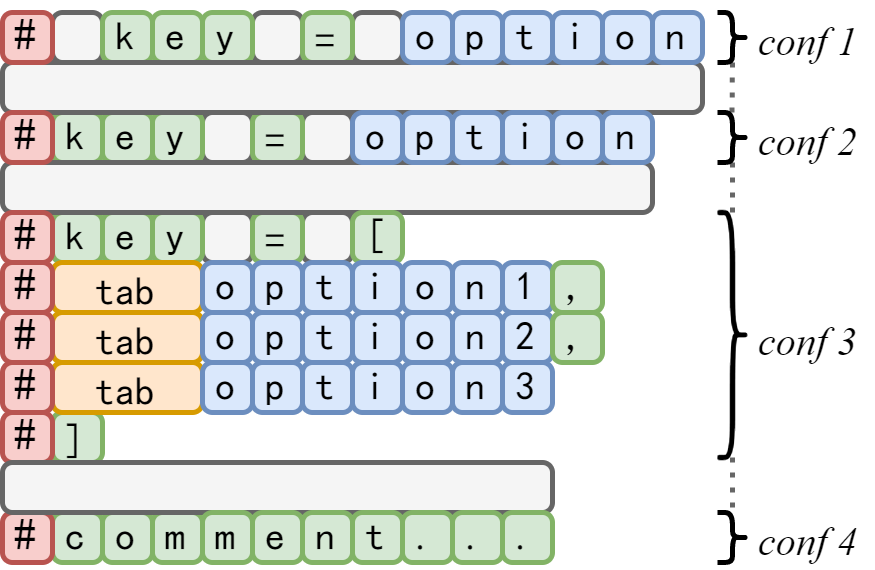}
	\caption{Different code formats of the configuration files.}
	\label{fig7}
\end{figure}

The process algorithm of the module $Confi$ is the other key core processing algorithm of Anywhere. The main function of module $Confi$ is to repetitively update the configuration items of the Scrapy that is generated by Anywhere. Due to differences in the user habits, the writing format in the configuration file of the Scrapy project will be different. As shown in Figure \ref{fig7}, the basic configuration format of the Scrapy setting file is following the key-value equation. The key part is corresponding to the name of the configuration item, and the value part is corresponding to the detailed content of the configuration item, which we defined as the option in Figure \ref{fig7} . There are four kinds of the main writing format of the configuration items in the $setting.py$ of the Scrapy project. For example, the $conf1$ represents the normal and recommended type of writing format in a Scrapy configuration file. The $conf2$ represents the situation that there is not spacing between the annotation symbol $\#$ and the key item. The $conf3$ represents the situation that the option item of the key-value structure of the configuration item is multiple lines.  The $conf4$ represents the situation that there is no configuration item in this line, and it is a comment, which will be filtered in the processing of module $Confi$.

Making clear about the writing format of configuration files in a Scrapy project, the next operation the Anywhere algorithm needs to implement is extracting the core information of a configuration code line from the $settings.py$ in a Scrapy project. As shown in Figure \ref{fig7}, there are some different kinds of writing of the configuration items we need to consider. The key core is to gain the location and detailed information of the key of the configuration in a line. Another key point needs to be noted is that for a code line that has not equal symbol $\#$, should not make the further processing. For those configuration items that are commented and consist of multiple lines, only needed to focus on the first line. For those new configuration items that have multiple sub items to configure, will be inserted in the $settings.py$ file at where the first lines of the same items were located repeatedly before.

\begin{figure}[t]
	\centering
	\includegraphics[width=0.9\columnwidth]{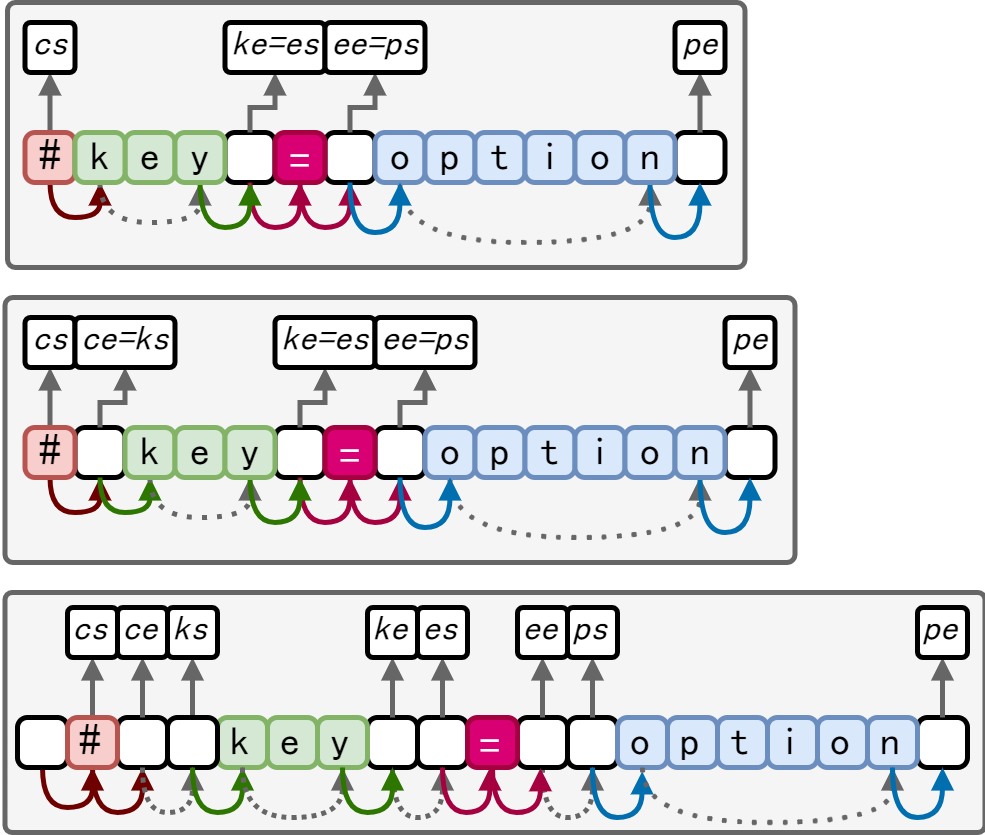}
	\caption{Key parameters extraction for the config-files.}
	\label{fig8}
\end{figure}

As shown in Figure ref{fig8}, we design a traversal judgement process with $for$ loop to gain the component of the key for the target configuration code line. However, the core processing is to gain the location of the target configuration code line, and the key component of one code line in the $settings.py$ is sole, therefore the core processing is to gain the expression of the key component between the indexes $ks$ and $ke$. The main method is to calculate and judge the null character bits between the comment symbol $\#$ and equation symbol $=$, followed by removing the spaces in the start and the end part to gain the key component. If there is not space in the start or end part, get the middle text between the comment symbol $\#$ and equation symbol $=$, which is used as the $key$ component. Then return the index of the configuration code line for further deleting and replacing operations. This process can be implemented multiple times during the design and programming of the Anywhere.

% <<:-----------------------------------------------------------------------------------

% >>:-----------------------------------------------------------------------------------
\subsection{Anywhere Framework Code Design Analysis}
\label{cm6}

In this subsection, we will analyse and discuss the key actual implementation code programming of the Anywhere framework to better understand the running and working principles of Anywhere.

\subsubsection{Key code for module Codein}

As shown in Code Listing 1, the function $codein\_block$ is the main function of the module $Codein$. The process of this is to insert the code into the file that is set by the parameter $file\_path$ at the location specified by the parameter $block_codes_list$. The parameter $block_code\_list$ stores the indexes and nest relationship information for the quickly locating of the target code line. The parameter $fb$ is used to set the insertion site of whether front or back. The main processing procedure is, first, get the insertion index with the function $code\_block\_ind$. Second, make the judgement of whether the returned index value is null. If the returned index is not null, that means the insertion location is valid and code can be inserted directly with the returned index that is gained from the analysis for the nested code blocks.

\begin{lstlisting}[language=Python, caption = Main function of module Codein]
def codein_block(file_path, block_codes_list, code, fb='b'):
    codes = jtc.readLines_file(file_path)
    ind = codein_block_ind(codes, block_codes_list)
    if ind != None:
        tab = tab_retract(codes)
        tabs = front_tab_nums(codes[ind-1])
        code = tab*' '*tabs + code
    elif len(block_codes_list) == 1:
        ind2 = code_inds_list(codes,
            block_codes_list[0])
        ind = ind2[0]
    if fb == 'b':
        codein_index(file_path, ind, code)
    elif fb == 'f':
        codein_index(file_path, ind-1, code)
\end{lstlisting}

\begin{lstlisting}[language=Python, caption = Gaining the corresponding index.]
def code_inds_list(codes, code):
    code_index_list = []
    for ind in range(len(codes)):
        if codes[ind].strip() == code:
            code_index_list.append(ind+1)
    return code_index_list
\end{lstlisting}

Otherwise, if the returned index is null, that means the insertion location can not be calculated and there is no nested code block in the $block\_codes\_list$. In this case, as shown in Code Listing 2, the method to get the parameter $ind$ is directly using the function $codein\_block\_ind$ to get the target index.

\subsubsection{Key code for module Codein}

As shown in Code Listing 3, the main function of module $Confi$ is the function $config\_option$. Amongst the import parameters, $config\_file\_path$ is used to indicate the path of the $settings.py$ configuration file of module $Confi$, $key$ is used to search the specific line of target configuration code line and $option$ is the corresponding value of the $key$. The parameter $sleep$ is used to set the sleep time in the checking existing procedure for the configuration file.

\begin{lstlisting}[language=Python, caption = Main function of module Confi.]
def config_option(config_file_path, key, option, sleep=0.1, equal='=', end='\n'):
    ...
    while True:
        time.sleep(sleep)
        if jtc.if_path_or_file_exist(config_file_path):
            break
    config_lines_list = jtc.readLines_file(config_file_path)
    ...
    new_lines = ''
    got_key = False
    
    for line in config_lines_list:
        if key in line and jtc.if_subStr_pureIn_str(equal, line) and end in line:
            got_key = True
            try:
                split_str_list = line.split(equal)
                key_ind = split_str_list[0].index(key)
                head_str = split_str_list[0][:key_ind]

                if option_str == '#':
                    if '#' in head_str:
                        head_str = ''
                    else:
                        head_str = '#' + head_str
                    option_str = split_str_list[1].strip()
                else:
                    head_str = ''
                new_line = head_str + key + ' ' + equal + ' ' + option_str
                new_lines += new_line + end
            except: pass
        else: new_lines += line
    if not got_key:
        new_lines += '\n' + key + ' ' + equal + ' ' + option_str + end
    jtc.write_file(config_file_path, new_lines)
\end{lstlisting}

As shown in Code Listing 3, if the configuration file exists, the first step is to get the detailed content of the configuration file and define the value $config\_lines\_list$ to store the data. Subsequently, using a for loop to make a traversal operation for each line item in the $config\_lines\_list$ to judge whether the key is in the current code line. The next step is judging whether the current code line is commented, if it is commented, deleting the comment symbol, which is represented from Line 22 to Line 26 and the result is the $key$ component string. Then generating an available expression of the $option$ component string. Finally, combining the $key$ and $option$ to generate a normal key-value configuration item expresion code line, followed by rewriting them into the configuration file.
% <<:-----------------------------------------------------------------------------------
% <<:=======================================================================================

% >>:=======================================================================================
\section{Experiment and Evaluation}
To evaluate the performance efficiency of the processing algorithm of Anywhere, we design the following corresponding test experiments. Our work is mainly focusing on improving the native Scrapy framework in quickly generating one or multiple Scarpy projects based on specific custom templates in the coding interaction level with the corresponding configuration changing in the meantime. Therefore, we mainly compared the Anywhere with the normal Scrapy framework in this task. We use the time of finishing in seconds to evaluate the speed and efficiency of the performances.

\begin{table}[htb]   
\begin{center}   
\caption{Experiments to test the performance of Anywhere.}  
\label{tb2} 
\tabcolsep=0.08cm
\renewcommand\arraystretch{1.5}
\begin{tabular}{|c|c|c|c|c|c|c|c|c|}   
\hline   \textbf{framework}  &  \textbf{task}  &  \textbf{config}  &  \textbf{time /s}  & \textbf{comparison} \\   

\hline Scrapy       &  Single Project        & No    & 4-9   & 100\%         \\
\hline Anywhere     &  Single Project        & No    & 2-5   &  200\%         \\
\hline Scrapy       &  Single Project        & Yes   & 6-12   & 100\%         \\
\hline Anywhere     &  Single Project        & Yes   & 3-6     & 200\%         \\

\hline Scrapy       &  Multiple Projects        & No    & 26-30   & 100\%         \\
\hline Anywhere     &  Multiple Projects        & No    & 9-10   & 300\%          \\
\hline Scrapy       &  Multiple Projects        & Yes   & 29-40 & 100\%           \\
\hline Anywhere     &  Multiple Projects        & Yes   & 10-13   & 300\%         \\

\hline   
\end{tabular}   
\end{center}   
\end{table}

Due to the individual difference of the testing user is big for they have different experiences of Scrapy and Anywhere, we make a big value interval in comparison part. As shown in Table \ref{tb2}, the count number of the multiple projects is 3. The value interval of comparison is 50\%. From the result we can see that the framework Anywhere can improve the generation and configuration efficiency of using Scrapy at a good level.

% <<:-----------------------------------------------------------------------------------
% <<:=======================================================================================

% >>:=======================================================================================
\section{Conclusion}
This paper presents a useful management toolkit for the single or multiple Scrapy project. The management for the multiple Scrapy projects is centralised. With Anywhere, improving the efficiency of generating Scrapy projects at the coding level will be easier. Moreover, Anywhere supports generating multiple Scrapy projects in batches with the preset code templates and repetitive configuration for the individual Scrapy projects at coding level. Anywhere will not change the code of Scrapy native framework but can improve the performance and efficiency of Scrapy native framework in quickly generating multiple projects with preset templates and automatically changing the configuration file of the specific Scrapy projects. With the relative experiments, the availability and high efficiency of Anywhere framework is verified.

There are still some shortcomings of the Anywhere framework, such as the lack of GUI operating support, the lack of the support of repetitive editing in module $Codein$, and weak in high concurrent processing. These shortcomings will be dealt with or solved by improving the design and processing algorithm code, which has a great research potential.
% <<:=======================================================================================

% \begin{acks}
% \end{acks}

\bibliographystyle{ACM-Reference-Format}
\bibliography{sample-sigconf}

% \appendix









\end{document}